\documentclass[10pt,onecolumn]{article} 

\usepackage[left=2cm, right=2cm, top=2cm]{geometry}
\usepackage{graphicx}
\usepackage{amssymb}
\usepackage{amsmath}
\usepackage[superscript]{cite}
\usepackage[detect-family = true,detect-inline-family = math, detect-weight = true,detect-inline-weight = math,separate-uncertainty=true, multi-part-units = single, product-units=single]{siunitx}
\usepackage{color}
\usepackage[utf8]{inputenc}
\usepackage{authblk}
\usepackage{lipsum}
\usepackage{hyperref}

\usepackage[doublespacing]{setspace}
\usepackage{caption}
\definecolor{mylblue}{rgb}{0.57, 0.73, 0.74}

\newcommand{\vv}{v}
\newcommand{\tq}{\tilde{q}}
\newcommand{\tl}{\tilde{\ell}}
\newcommand{\tv}{\tilde{\vv}}
\newcommand{\sech}{\mathrm{sech}}

\title{Splitting and recombination of bright-solitary-matter waves}

\author[1]{Oliver J. Wales}
\author[1]{Ana Rakonjac}
\author[2]{Thomas P. Billam}
\author[3]{John L. Helm}
\author[1]{Simon A. Gardiner}
\author[1]{Simon L. Cornish}
\affil[1]{Joint Quantum Centre (JQC) Durham--Newcastle, Department of Physics, Durham University, South Road, Durham DH1 3LE, UK}
\affil[2]{Joint Quantum Centre (JQC) Durham--Newcastle, School of Mathematics, Statistics and Physics, Newcastle University, Newcastle upon Tyne NE1 7RU, UK}
\affil[3]{Dodd-Walls Centre for Photonic and Quantum Technologies, Department of Physics, University of Otago, 730 Cumberland Street, Dunedin 9016, New Zealand}

\date{\today}
\begin{document}
\maketitle
\begin{abstract}
    Solitons are long-lived wavepackets that propagate without dispersion and exist in a wide range of one-dimensional (1D) nonlinear systems. A Bose-Einstein condensate trapped in a quasi-1D waveguide can support bright-solitary-matter waves (3D analogues of solitons) when interatomic interactions are sufficiently attractive that they cancel dispersion. Solitary-matter waves are excellent candidates for a new generation of highly sensitive interferometers, as their non-dispersive nature allows them to acquire phase shifts for longer times than conventional matter-waves interferometers. However, such an interferometer is yet to be realised experimentally. In this work, we demonstrate the splitting and recombination of a bright-solitary-matter wave on a narrow repulsive barrier, which brings together the fundamental components of an interferometer. We show that both interference-mediated recombination and classical velocity filtering effects are important, but for a sufficiently narrow barrier interference-mediated recombination can dominate. We reveal the extreme sensitivity of interference-mediated recombination to the experimental parameters, highlighting the potential of soliton interferometry.  
\end{abstract}

Bright-solitary waves, referred to as solitons in this work, are wavepackets that propagate in a quasi-1D geometry without dispersion, owing to a self-focussing nonlinearity. They are of fundamental interest in a broad range of settings due to their ubiquity in nonlinear systems, which occur prolifically in nature \cite{dynamics_nearly_integrable,solitons_nonlinear_lattice}{.} In Bose-Einstein condensates (BECs) the nonlinearity is provided by interatomic interactions governed by the $s$-wave scattering length, which can be tuned using a magnetic Feshbach resonance \cite{feshbach_review}{.} Bright solitons in BECs of $^7$Li, $^{85}$Rb, $^{39}$K and $^{133}$Cs have so far been experimentally demonstrated \cite{soliton_formation_hulet,soliton_formation_salomon,soliton_formation_cornish,controlled_formation,solitonic_interferometer,Lepoutre2016Soliton,Cs_soliton}{.} Understanding and probing the coherent phase carried by matter-wave solitons is an area of particular relevance for BEC physics, both because it is important in determining the stability of soliton-soliton collisions \cite{soliton_collisions_hulet,Cs_soliton,collisions_solitons,realizing_soliton_phase,Condensate_soliton_collisions} and because there is a great interest in using solitons for atom interferometry \cite{quantum_dynamics_interferometer,interferometer_nonlinear_splitter,dipolar_splitting,soliton_interferometer,sagnac_interferometer,splitting_solitons,soliton_collisions_barrier,soliton_interactions_barrier}{.} 

Matter-wave interferometers have emerged as a means of achieving unprecedented sensitivity in interferometric measurements \cite{interferometry_atoms_molecules,adams1994atom,godun2001prospects,pritchard2001atom}{.} However, they have typically been limited by either interatomic collisions or dispersion of the atomic wavepackets, which cause dephasing and a reduced signal to noise, respectively \cite{Integrated_BEC_interferometer}{.} Previous works have successfully reduced the impact of interatomic collisions through the control of interatomic interactions \cite{weakly-interacting_interferometer,dephasing_bloch}{,} or by generating squeezed states \cite{spin_squeezing,number_squeezing}{.} However, dispersion remains a limitation. A soliton-based interferometer has the potential to overcome dispersion, allowing for much longer phase-accumulation times, albeit for an increased quantum noise \cite{Haine_2018}{.} To date, only one experiment has demonstrated interferometry with a soliton \cite{solitonic_interferometer}{,} in which Bragg pulses were used for splitting and recombination. However, interferometer times were insufficient to exploit the non-dispersive property of solitons. 

\begin{figure}[htbp]
\centering
\def\svgwidth{0.475\linewidth}
\graphicspath{{Figures/}}
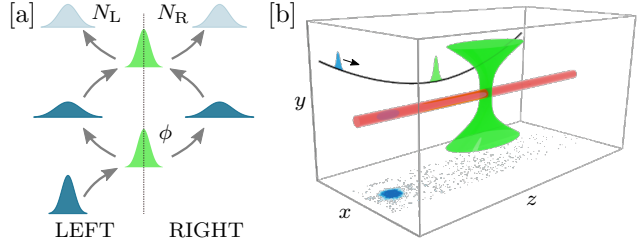
\caption{A soliton interferometery scheme based on a repulsive Gaussian barrier. [a] The initial soliton (blue) is split into two daughter solitons by the barrier (green). They return to the barrier and interfere, with resultant population fractions on the left and right ($N_\mathrm{L}$ and $N_\mathrm{R}$ respectively) determined by the phase difference ($\phi$) between the daughter solitons. [b] A sketch of the experimental implementation of the barrier-based interferometer. The 3D plot and $yz$-plane highlight the optical waveguide (red) and the magnetic harmonic potential along the waveguide, respectively, with the barrier shown in green for both. The $xz$-plane shows an example image of a soliton in this potential.}
\label{schematic}
\end{figure}

Narrow repulsive barriers have been proposed as atomic beam splitters for soliton-based interferometers \cite{interferometer_nonlinear_splitter,dipolar_splitting,soliton_interferometer,sagnac_interferometer,quantum_dynamics_interferometer,soliton_collisions_barrier,splitting_solitons,soliton_interactions_barrier,soliton_splitting_tunneling,soliton_hong-ou-mandel_theory}{,} as illustrated in Fig.~\ref{schematic}. At the Gross-Pitaevskii level, a soliton incident on such a barrier is split cleanly into transmitted and reflected daughter solitons, provided the incident velocity is sufficiently fast that the effects of interatomic interactions can be neglected during the splitting \cite{Holmer2007}{.} When these daughter solitons are subsequently made to spatio-temporally overlap at the barrier, total or partial interference-mediated recombination occurs, depending on their relative phase \cite{soliton_collisions_barrier}{.} This phase-sensitive splitting and recombination forms the basis of the interferometer. 

In this work, we demonstrate the splitting and recombination of a soliton on such a repulsive Gaussian barrier. We show that for a barrier much wider than the soliton width, classical velocity filtering effects dominate, precluding applications to interferometry. In this case, we observe that the majority of the population consistently appears on the original \textit{left} side of the barrier after the second barrier interaction. However, for a barrier width approaching the soliton length-scale, interference becomes significant, despite velocity filtering effects still being present and measurable. In this case, the majority of the population can appear on \textit{either} side of the barrier following recombination, varying from shot-to-shot due to the extreme sensitivity of interference-mediated recombination to experimental parameters. Our results show good qualitative agreement with Gross-Pitaevskii equation (GPE) simulations throughout. 

\subsection*{Results}
\subsubsection*{Controllable splitting}

We form a soliton of approximately 2500 $^{85}$Rb atoms in a quasi-1D waveguide, following procedures described elsewhere \cite{magnetic_transport,quantum_reflection,direct_evaporation,controlled_formation}{.} An additional harmonic magnetic potential produces axial trapping frequencies of up to $\SI{1.5}{Hz}$ along the waveguide. The soliton undergoes centre of mass oscillations of controllable amplitude in this potential. Measurements are taken using a destructive absorption imaging technique (see Methods); throughout this paper, each image represents an individual run of the experiment. We have observed unprecedented soliton lifetimes of longer than $\SI{20}{s}$ in this potential, corresponding to $\sim30$ centre of mass oscillations and a total distance covered of over $\SI{2}{cm}$ (O.J.W., A.R. and S.L.C., in preparation).

\begin{figure}[htbp]
\centering
\def\svgwidth{0.475\linewidth}
\graphicspath{{Figures/}}
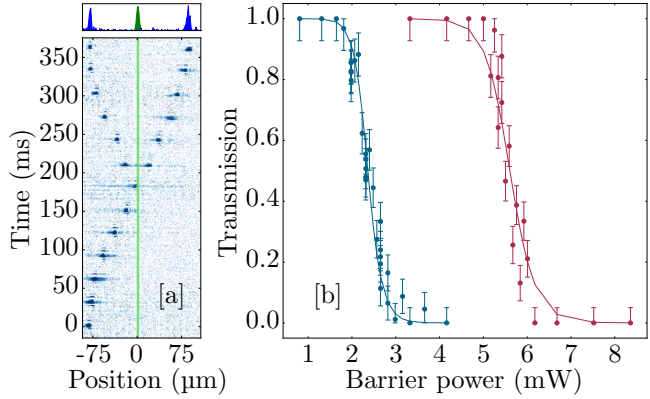
\caption{Controllable splitting of a soliton into two daughter solitons by a repulsive Gaussian barrier. [a] An example sequence of a soliton being split by the narrow barrier into two solitons of approximately equal atom number. The upper panel shows the column density across the final image in the sequence. [b] The transmission of a slow (blue) and fast (red) soliton through the wide barrier as the barrier power is varied. The solid lines are quasi-1D GPE simulations that are fit to the data by varying the barrier width.}
\label{splitting_figure}
\end{figure}

To split a soliton, we use a repulsive Gaussian barrier formed by a blue-detuned highly-elliptical laser beam which is focussed down to bisect the waveguide (Fig.~\ref{schematic}). We investigate two barrier widths: a ``wide'' barrier and a ``narrow'' barrier, with experimentally-determined waists ($1/$e$^2$ radii) of $10.6~^{+0.5}_{-0.1}~\si{\micro m}$ and $\SI{3.6\pm 0.4}{\micro m}$, respectively (see Methods). Upon reaching the barrier, the soliton is either reflected, transmitted, or split into two daughter solitons (Fig.~\ref{splitting_figure}). The barrier height and therefore the transmission through the barrier is tuned experimentally by varying the total barrier power. 

The nature of the splitting mechanism depends critically on the barrier width. In the limit of a $\delta$-function barrier, quantum tunneling dominates and the \textit{area} of the barrier potential determines the transmission probability \cite{splitting_solitons}{.} However, for barriers wider than the soliton width, the transmission probability instead depends primarily on the incident soliton's centre of mass kinetic energy relative to the barrier \textit{height}. The transmission probability is well-approximated by the analytic result for a $\mathrm{sech}^2$ potential \cite{landau_quantum} (see Methods), which becomes a step function in the classical limit of an infinitely wide barrier. 

For the wide barrier, quasi-1D GPE simulations yield $\SI{50}{\%}$ transmission when the barrier height is only $\SI{1}{\%}$ higher than the kinetic energy of the soliton at the barrier, implying that the splitting mechanism is almost entirely a classical process. However, equivalent simulations for the narrow barrier yield a barrier height that is $\SI{11}{\%}$ higher than the kinetic energy, suggesting that quantum tunneling plays a small role. Experimentally, we measure $\SI{50}{\%}$ transmission when the barrier height is $\SI{11\pm2}{\%}$ and $\SI{35\pm16}{\%}$ higher than the soliton kinetic energy at the barrier, for the wide and narrow barriers respectively. Our measurements verify that quantum tunneling is more relevant for the narrow barrier than the wide barrier, though classical splitting is the dominant effect in both cases. 

Interestingly, the steep energy dependence of the measured transmission functions produces a velocity filtering effect, whereby the transmitted soliton always has a higher centre of mass kinetic energy than the reflected soliton \cite{soliton_interferometer,soliton_interactions_barrier,soliton_splitting_tunneling}{.} This causes the transmitted soliton to have a larger oscillation amplitude than the reflected soliton, which we directly observe in the trajectories of the daughter solitons (see Extended Data Fig.~\ref{velocity_selection}). 

\subsubsection*{Recombination}

Interference-mediated recombination occurs when the two daughter solitons return to the barrier and spatio-temporally overlap. Following this second barrier interaction, the resultant populations on each side of the barrier are determined by the relative phase between the daughter solitons. It is important to note that the barrier is integral to interference-mediated recombination; in the absence of the barrier, the two daughter solitons simply pass through one another (see Extended data Fig.~\ref{asymmetric_splitting}). 

In the limit of a $\delta$-function barrier, theoretical studies of soliton splitting indicate that there is an intrinsic $\pi/2$ phase difference between the daughter solitons \cite{soliton_collisions_barrier}{.} In our harmonic potential, it is expected that this phase difference is maintained and that we should ideally achieve completely constructive (destructive) interference on the right (left) of the barrier, resulting in a fully recombined soliton appearing on the \textit{right}. However, velocity filtering confounds this ideal outcome. If we remove the effects of interference and consider velocity filtering alone, the reflected (transmitted) daughter soliton is always primarily reflected from (transmitted through) the barrier at the second barrier interaction, resulting in a single soliton appearing on the \textit{left}. This should be considered a \textit{merging} of the two daughter solitons, rather than true \textit{recombination}, as it is mediated by classical velocity filtering and not interference.

\begin{figure}[htbp]
\centering
\def\svgwidth{0.475\linewidth}
\graphicspath{{Figures/}}
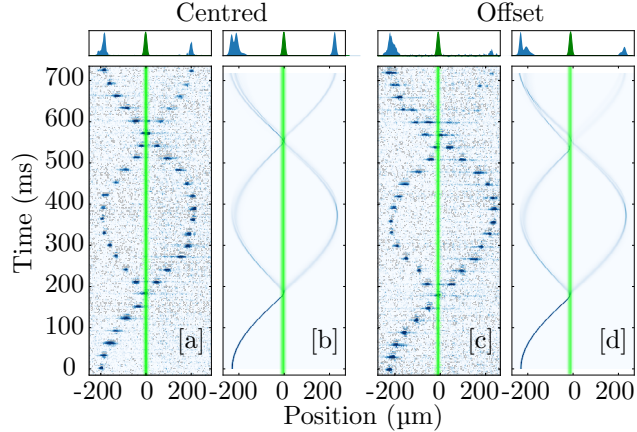
\caption{Trajectories of a soliton undergoing centre of mass oscillations and interacting twice with the wide barrier. The barrier power is held constant throughout to give $\SI{50}{\%}$ transmission at the first barrier interaction. [a] and [c] show experimental image sequences, for a centred barrier and a barrier offset by $\SI{10}{\micro m}$, respectively. The upper panels are column densities of the final images in the sequence, with the barrier shown in green. [b] and [d] are quasi-1D GPE simulations of [a] and [c], respectively}
\label{wide_barrier}
\end{figure}

\begin{figure}[htbp]
\centering
\def\svgwidth{0.475\linewidth}
\graphicspath{{Figures/}}
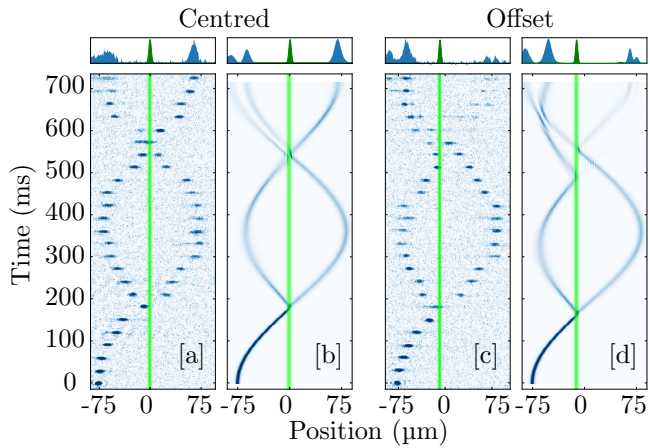
\caption{Trajectories of a soliton undergoing centre of mass oscillations and interacting twice with the narrow barrier. The barrier power is held constant throughout to give $\SI{50}{\%}$ transmission at the first barrier interaction. [a] and [c] show experimental image sequences for a centred barrier and a barrier offset by $\SI{10}{\micro m}$, respectively. The upper panels are column densities of the final images in the sequence, with the barrier shown in green. [b] and [d] are quasi-1D GPE simulations of [a] and [c], respectively.}
\label{narrow_barrier}
\end{figure}

To isolate and expose the effects of velocity filtering experimentally, the barrier is offset from the centre of the harmonic potential, preventing the daughter solitons from spatio-temporally overlapping during the second barrier interaction (Fig.~\ref{wide_barrier}[c] and Fig.~\ref{narrow_barrier}[c]). This prevents any possibility of interference. We observe that the population appears almost entirely on the left after the second barrier interaction, which is consistent with strong velocity filtering. In reality, neither complete interference nor total velocity filtering can be achieved, as it is impossible to realise a $\delta$-function barrier and interactions preclude total velocity filtering (see Methods). 

To study interference-mediated recombination, the barrier is aligned with the centre of the harmonic potential to ensure maximal spatio-temporal overlap of the daughter solitons with the barrier. For the wide barrier, we observe that the majority of the population still appears on the \textit{left} after the second barrier interaction (Fig.~\ref{wide_barrier}[a]), suggesting that velocity filtering dominates the recombination process. 

The outcome for the narrow barrier is markedly different, as shown in Fig.~\ref{narrow_barrier}[a]. In this case, we can clearly see that the majority of the population is able to finish on the \textit{right}, which can \textit{only} be explained by interference-mediated recombination. This occurs despite the presence of measurable velocity filtering (see Fig.~\ref{narrow_barrier}[c] and Extended data Fig.~\ref{velocity_selection}). The final five images in Fig.~\ref{narrow_barrier}[a] are post-selected to show the largest observed population on the right in order to facilitate a direct comparison with the quasi-1D GPE simulations. In practise, the extreme sensitivity of interference-mediated recombination to the experimental parameters leads to large shot-to-shot fluctuations in the final populations on the left and right, as discussed in detail below.

\begin{figure}[htbp]
\centering
\def\svgwidth{0.475\linewidth}
\graphicspath{{Figures/}}
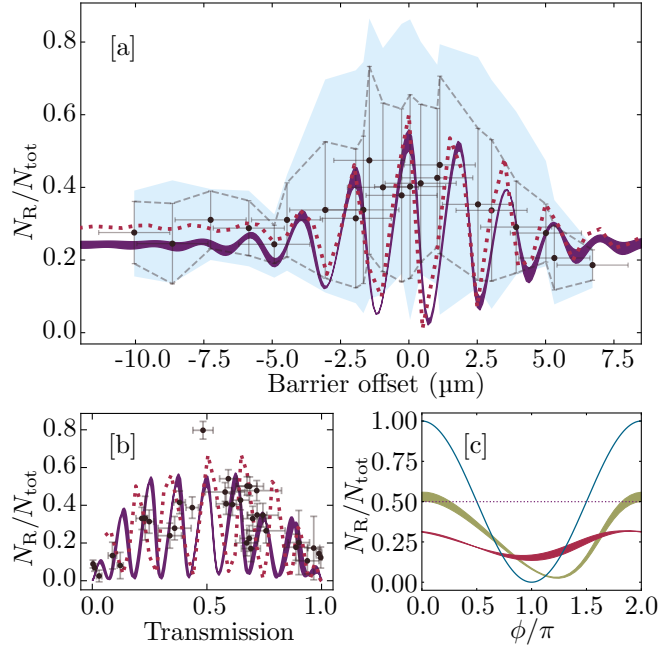
\caption{The fraction of atoms on the right of the narrow barrier following the second barrier interaction, as a function of barrier offset [a], barrier transmission for a centered barrier [b] and daughter soliton phase difference [c]. In [a], the data points show the mean and the dashed grey envelope (blue shaded region) indicates the standard deviation (maximum and minimum values) across the 5--10 measurements taken at each offset position. [b] For a range of barrier heights, images were taken before and after the second barrier interaction in order to determine the transmission and its uncertainty. The error bars in $N_\mathrm{R}/N_\mathrm{tot}$ arise from uncertainty in the fit to determine the number of atoms. The solid purple regions (red dotted lines) in both [a] and [b] are quasi-1D (3D) GPE simulations. [c] The theoretically predicted interference effect for coherent daughter solitons with a controllable relative phase ($\phi$). The red (green) curves show predictions for the wide (narrow) barrier, the solid blue line is for an idealised $\delta$-function barrier at high velocity and the horizontal dotted line is for a $\delta$-function barrier with non-interfering daughter solitons.}
\label{recombination}
\end{figure}

We explore the dependence of interference-mediated recombination on the offset of the narrow barrier in Fig.~\ref{recombination}. Theoretically, we observe oscillations in the fraction of atoms on the right of the barrier as the barrier offset is varied. Here, the barrier offset introduces a position shift of the transmitted and reflected wavepackets, which in turn leads to velocity-induced phase gradients across the wavepackets when they recombine. This results in the observed interference fringes. These fringes are modulated by an envelope caused by the changing spatio-temporal overlap between the wavepackets (see Methods). Experimentally, we observe an increased shot-to-shot fluctuation when the barrier is closer to the centre of the harmonic potential within an envelope that is in good qualitative agreement with those predicted by quasi-1D and 3D GPE simulations (see also Extended data Fig.~\ref{standard_deviation}). The predicted oscillatory behaviour is not resolved because the shot-to-shot variation of the axial harmonic potential relative to the barrier position is larger than the fringe spacing (see Methods). Note that the theory lines in Fig.~\ref{recombination} have no free fitting parameters and the experimental values for the barrier offset are determined from independent measurements of the trap and barrier positions. 

We additionally measure the final populations as a function of barrier transmission for the centred narrow barrier (Fig.~\ref{recombination}[b]). Here, the fringes seen in the theory arise from the differing number-dependent chemical potentials \cite{BEC_theory_review} between the daughter solitons, inducing a phase-winding effect. This is modulated by an envelope, caused by the varying spatial overlap of the daughter soliton wavepackets as one becomes larger than the other. We again observe good qualitative agreement with the theory, with a higher population on the right after the second barrier interaction for $\SI{50}{\%}$ transmission than for the transmission extremes.

\subsection*{Discussion}

The experimental observations of Figs.~\ref{narrow_barrier} and \ref{recombination} are definitive signatures of interference-mediated recombination, as the strong increase in population on the right of the barrier and the increase in population fluctuations can only be explained by interference effects. Without interference, there is no significance to the increased spatio-temporal overlap of the daughter solitons when the barrier is centred, so the same populations and fluctuations would be recovered for all offsets. To an extent, this is seen between the centred and offset cases for the wide barrier in Fig.~\ref{wide_barrier}. An additional artefact of interference is seen in the double-soliton structure on the left of the barrier following the second barrier interaction in Fig.~\ref{narrow_barrier}[b]; further theoretical simulations demonstrate that introducing an additional phase difference between the daughter solitons can merge the two solitons into one, leaving only a small population on the right of the barrier. This phase dependence is explored theoretically in Fig.~\ref{recombination}[c]; changes in phase difference between the daughter solitons result small variations in the final populations for the wide barrier, whereas they result in far more significant variations for the narrow barrier. 

Our experimental results can be understood in terms of relatively simple analytic estimates for the populations after the second barrier interaction (see Methods). In particular, the spacing of fringes as a function of barrier offset depends on the soliton velocity and the atomic mass $m$, while the width of the envelope depends on the soliton width $\ell_\mathrm{s} = \hbar / (2 m \omega_\mathrm{r} |a_\mathrm{s}| N)$, where $\omega_\mathrm{r}$ is the radial trap frequency, $a_\mathrm{s}$ is the scattering length, and $N$ is the atom number. The analytic model provides good estimates when the velocity of the soliton at the barrier $v \gtrsim 2 \omega_\mathrm{r} |a_\mathrm{s}| N$; slower velocities ($v \ll 2 \omega_\mathrm{r} |a_\mathrm{s}| N$) generate much more complex interference fringes due to the nonlinear interactions, which are more significant during longer recombination times.

It is clear that exceptional stability in the relative position between the barrier and axial harmonic potential is required to create a viable interferometer; to resolve the oscillatory behaviour in Fig.~\ref{recombination} requires the barrier to be controllable at the level of $\sim\SI{0.1}{\micro m}$ with respect to the harmonic potential. Currently, the axial potential in our experiment is generated magnetically, making it susceptible to the ambient magnetic field. Remarkably, a shot-to-shot variation of only $\SI{3}{mG}$ is sufficient to fully account for the observed fluctuations. This sensitivity to magnetic field could be removed altogether using an all-optical potential generated by, for example, acousto-optic deflectors \cite{arbitrary_potential_Boshier,arbitrary_potential_Rubinsztein-Dunlop} or a digital micromirror device \cite{dmd_neely}{.} Optical methods are also experimentally attractive because the shot-to-shot stability of our current optical potentials is $\sim\SI{0.3}{\micro m}$ and would probably be sufficient to observe the oscillatory behaviour in Fig.~\ref{recombination}{.} Furthermore, these methods offer the flexibility for more complicated geometries, such as a ring-shaped trap for a soliton Sagnac interferometer \cite{sagnac_interferometer}{.} Alternatively, other atomic systems may prove to be more resilient to barrier position. For instance, the lower mass of lithium would result in broader fringes, leading to less-stringent requirements of the relative position stability. The lower mass has the added benefit of making the solitons themselves larger, so the barrier is comparatively narrower and velocity filtering is suppressed. 

Our apparatus also lends itself to soliton-soliton collision experiments. As the relative phase of the daughter solitons is expected to be well-defined and controllable, the outcome of soliton-soliton collisions could be controlled completely deterministically, unlike in other reported experiments \cite{Cs_soliton,soliton_collisions_hulet}{.} The ability to manipulate the daughter solitons' relative velocity and population fractions also allows us to access a wide parameter range of interest \cite{realizing_soliton_phase,collisions_solitons}{.} We have observed the daughter solitons to undergo many ($>10$) soliton-soliton collisions in the absence of the barrier without any instances of mergers or collapse. This is a strong experimental marker for long coherence times. However, we cannot yet conclusively determine whether the splitting process truly retains coherence between the daughter solitons, as the shot-to-shot fluctuations seen in Fig.~\ref{recombination} could be equally explained by phase noise (see Extended data Fig.~\ref{standard_deviation}). A measurement of either the fringes seen in Fig.~\ref{recombination} or controllable, deterministic soliton-soliton collisions would be definitive proof of coherence between the daughter solitons.

\subsection*{Conclusion}

Our measurements are the first conclusive realisation of splitting and interference-mediated recombination of bright-matter-wave solitons on a repulsive barrier. We have demonstrated the controlled splitting of a soliton into two daughter solitons, in good agreement with GPE simulations. We have shown that classical velocity filtering dominates interference during the recombination process for wider barriers, resulting in a \textit{merging} of the daughter solitons. However, with a reduced barrier width, interference overcomes velocity filtering effects and interference-mediated \textit{recombination} is observed. We have shown the extreme sensitivity of this setup to experimental parameters, by varying both the barrier transmission and the barrier offset. Whilst the required stability to exploit this intrinsic sensitivity is beyond the current configuration of the experiment, future realisations with reduced instabilities, all-optical potentials, or more favourable atomic species may be able to harness this scheme to further the precision of interferometric measurements and to gain a deeper understanding of the fundamental properties of matter-wave solitons.

\newpage
\subsection*{Methods}

\subsubsection*{Soliton production}

We create solitons of $\sim$2500 $^{85}$Rb atoms using methods described in previous publications \cite{direct_evaporation,quantum_reflection,magnetic_transport,controlled_formation}{.} Briefly, a BEC of up to $7000$ $^{85}$Rb atoms with a condensate fraction of $>\SI{80}{\%}$ is formed in a hybrid trap comprised of a red-detuned crossed optical dipole trap and a magnetic potential provided by quadupole and bias fields. The $s$-wave scattering length is set to $\sim\SI{220}{a_0}$ by tuning the magnetic field near the $\SI{165.75}{G}$ zero crossing of a broad magnetic Feshbach resonance in the $F=2$, $m_\mathrm{F}=-2$ state \cite{feshbach_resonance_durham,feshbach_resonance_wieman,feshbach_resonance_wieman2}{.} To form a soliton, the scattering length is first ramped to zero over $\SI{100}{ms}$ before simultaneously removing one dipole trapping beam and jumping the scattering length to a small negative value. Stable solitons are formed in the resulting waveguide for scattering lengths in the range of $\SI{-13.5}{a_0}$ $\lesssim$ $a_\mathrm{s}$ $\lesssim$ $\SI{-7}{a_0}$, with soliton production possible up to $\sim\SI{-20}{a_0}$ for a reduced number of atoms (O.J.W., A.R. and S.L.C., in preparation).

An additional pair of magnetic coils produces a harmonic confining potential in the axial direction of the waveguide, with trapping frequencies of up to $\omega_\mathrm{z}/2\pi \sim \SI{1.5}{Hz}$, allowing us to observe axial centre-of-mass oscillations of the soliton along the waveguide. The centre of the harmonic potential is maneuvered along the axial direction using a pair of ``shim'' coils which, coupled with independent control over the barrier position, provides control of the soliton velocity at the barrier. As the harmonic potential is magnetic, the soliton experiences a changing magnetic field as it oscillates. However, the field varies by $<\SI{1}{mG}$ across a typical centre of mass oscillation. This has a negligible effect on the scattering length ($<0.1a_0$ at $a_\mathrm{s}=\SI{-10}{a_0}$) and so does not effect the stability of the soliton.

\subsubsection*{Soliton beam-splitter}

The light for the Gaussian barrier is produced by a $\SI{532}{nm}$ laser. A cylindrical lens forms a highly elliptical beam, which is focussed onto the waveguide such that the narrow axis is oriented along the axial direction of the trap (see Fig.~\ref{schematic}[c]). The barrier potential height is controlled by changing the beam power and the barrier position along the waveguide can be precisely adjusted via a piezo-actuated mirror. The two barrier widths investigated in this work are generated using two different objective lenses, with focal lengths of $\SI{100}{mm}$ and $\SI{30}{mm}$ for the wide and narrow barriers, respectively. 

To characterise the width of the narrow barrier, the beam is exposed onto an elongated cloud of thermal atoms and the resultant dip in atomic density observed in the images is fitted using a Gaussian profile (see Extended Data Fig.~\ref{cloud_depletion}). Using this technique, we measure a width along the waveguide axial direction of $w_\mathrm{z}=\SI{4.7\pm0.3}{\micro\metre}$ which, when corrected for the \SI{3.0\pm0.3}{\micro\metre} resolution limit of the imaging system (see below), becomes \SI{3.6\pm0.4}{\micro\metre}. By translating the thermal cloud across the barrier beam, we also determine the transverse width to be $w_\mathrm{x}=\SI{117\pm9}{\micro m}$ in the plane of the atoms.

To determine the width of the wide barrier, the beam is profiled outside of the vacuum chamber using a duplicate optical setup. Using this method, we measure an axial waist of $w_\mathrm{z}=10.6~^{+0.5}_{-0.1}~\si{\micro m}$ and a transverse width at the axial waist position of $w_\mathrm{x}=\SI{434\pm5}{\micro m}$. The asymmetry in uncertainty for $w_\mathrm{z}$ accounts for uncertainties in the position of the focus inside the vacuum chamber, which can only lead to a larger barrier width.

Quasi-1D GPE theory is also fitted to splitting data with the barrier width in the axial direction of the waveguide as the only free parameter (as in Fig.~\ref{splitting_figure}) for both the narrow and wide barriers. The transverse barrier widths are constrained by the experimental values above. Using this technique, we determine barrier widths of $w_\mathrm{z}=\SI{4.8\pm0.2}{\micro\metre}$ and $w_\mathrm{z}=\SI{11.9\pm0.3}{\micro\metre}$ for the narrow and wide barrier beams, respectively. The uncertainty in each value is the standard error in the fitted widths across several sets of splitting data, taken for various soliton velocities. Any assumptions of the quasi-1D GPE model that are not fulfilled in the experiment (for example in the transverse mode profiles) could contribute to the small discrepancy with the measured values.

\subsubsection*{Imaging}

We perform \textit{in-situ}, high-field, high-intensity absorption imaging of the solitons (O.J.W., A.R. and S.L.C., in preparation). Imaging at high field ensures that the soliton wavepackets are not perturbed by crossing the Feshbach resonance during trap turn-off, which broadens their shape significantly. Intense, short probe pulses are used to minimise width broadening as a result of photon recoil \cite{high_intensity,strong_saturation_absorption}{.}

At a magnetic field of $\sim\SI{165.85}{G}$, where experiments are performed, there are no closed optical transitions from the $F=2$, $m_\mathrm{F}=-2$ ground state. This is detrimental to imaging, as each atom can only scatter an average of $3.28$ photons before being lost to a dark state. We circumvent this by transferring atoms from $F=2$, $m_\mathrm{F}=-2$ to $F=3$, $m_\mathrm{F}=-3$ via microwave adiabatic rapid passage (ARP), from which a closed $\sigma^+$ transition exists. A typical imaging sequence begins with a $\SI{100}{\micro s}$, $\SI{300}{kHz}$ ARP sweep to transfer $\sim\SI{90}{\%}$ of the atoms to the imaging state, followed by a $\SI{10}{\micro\second}$ probe pulse with an intensity of $\sim\SI{10}{\mathrm{I_{sat}}}$. 

The resolution limit of the imaging system is expected to be of the order of the soliton width for the $\SI{30}{mm}$ objective lens system. The observed soliton in the image plane is the convolution of the soliton in the object plane with the point spread function of the imaging system. As the width of a soliton doubles when the atom number is halved, for a constant interaction strength, we can estimate the resolution limit of the imaging system by imaging the soliton before and after splitting, with the barrier set to $\SI{50}{\%}$ transmission. In this case we find that the resolution limit, $r$, is given by
\begin{equation}
    r = \sqrt{\frac{4k_N^2-k_{N/2}^2}{3}},
\label{resolution_equation}
\end{equation}
where $k_N$ and $k_{N/2}$ are the measured widths before and after splitting respectively. Over ten experimental runs each, we measure an average soliton width of \SI{3.6\pm0.2}{\micro\metre} and average width of the daughter solitons of $\SI{4.9\pm0.3}{\micro\metre}$, where the quoted uncertainty represents the standard error across the runs. This implies a resolution limit of \SI{3.0\pm0.3}{\micro\metre}, in good agreement with the diffraction-limited value of \SI{2.9}{\micro m} for our optical system. 

To account for the resolution limit in the GPE simulations, a $\SI{3}{\micro m}$ Gaussian convolution is applied to the quasi-1D GPE density profiles in Fig.~\ref{narrow_barrier}. As both the initial and daughter soliton widths are below the resolution limit for the $\SI{100}{mm}$ lens, it is impossible to accurately apply Eq.~\ref{resolution_equation}. Instead, the resolution limit is extrapolated from the measured $\SI{30}{mm}$ lens resolution using the ratio of the focal lengths of the two objective lenses. Therefore, a $\SI{10}{\micro m}$ Gaussian convolution is applied to the quasi-1D GPE density profiles in Fig.~\ref{wide_barrier}.

The change of objective lens also alters the camera's field of view. Therefore, it was necessary to reduce the centre of mass oscillation amplitude in Figs.~\ref{narrow_barrier} and~\ref{recombination} ($\SI{75}{\micro m}$) from that in Fig.~\ref{wide_barrier} ($\SI{225}{\micro m}$) to compensate.

\subsubsection*{Trap stability}

We determine the RMS shot-to-shot fluctuation in the position of the centre of the axial potential to set the offset uncertainty in Fig.~\ref{recombination}. This is found by taking ten repeat measurements of the soliton position both immediately after release and  after half a trap period, finding shot-to-shot variations of $\SI{0.3\pm0.1}{\micro m}$ and $\SI{2.6\pm0.6}{\micro m}$ respectively. These variations imply fluctuations in the position of the centre of the harmonic potential of $\SI{1.3\pm 0.3}{\micro m}$. This is comparable to the fringe period in Fig.~\ref{recombination} predicted by GPE simulations, meaning that the experiment samples some region of the fringes on each run, hence the increased variation when the barrier is close to the center of the harmonic potential. As the uncertainty is dominated by position fluctuations after a half trapping period, we attribute it to be dominated by magnetic potential instability. A stray field of only $\sim\SI{3.0}{mG}$ along the axial direction would account for this shot-to-shot fluctuation. From another set of ten sequences, we determine the barrier position to fluctuate from shot-to-shot with an RMS of $\SI{0.3 \pm 0.1}{\micro\metre}$. This is small enough to resolve the fringes in Fig.~\ref{recombination}, for a similarly-stable harmonic potential. 

\subsubsection*{Gross-Pitaevskii model}

Assuming a mean-field description, the collective wavefunction $\psi$ (normalized to unity) obeys the GPE
\begin{equation}
i \hbar \frac{\partial \psi}{\partial t} = \left[ \frac{-\hbar^2}{2m} \nabla^2 + \frac{m [\omega_\mathrm{z}^2 z^2 + \omega_\mathrm{r}^2 (x^2 + y^2)]}{2} + \mathcal{V}(x,z) + g_\mathrm{3D} |\psi|^2 \right] \psi, \label{eqn:3dgpe}
\end{equation}
where $m$ is the atomic mass, $\omega_\mathrm{z}$ is the axial trap frequency, $\omega_\mathrm{r}$ is the radial trap frequency, $g_\mathrm{3D} = 4 \pi \hbar^2 N a_\mathrm{s} / m$ where $a_\mathrm{s}$ is the $s$-wave scattering length and $N$ is the atom number. The optical barrier potential is modelled as $\mathcal{V}(x,z) = U(x) V(z)$, with
\begin{align}
U(x) &= \exp\left[\frac{-2 x^2}{w_\mathrm{x}^2}\right],\\
V(z) &= \frac{\alpha P}{\epsilon_0  c  \pi  w_\mathrm{z}  w_\mathrm{x}} \exp\left[\frac{-2 (z-z_\mathrm{off})^2}{w_\mathrm{z}^2}\right],
\end{align}
where $\alpha$ is the relevant polarisability for $^{85}$Rb, $z_\mathrm{off}$ is the barrier offset from the centre of the trap and $P$ is the optical power in the beam. 

Since $\omega_\mathrm{r} \gg \omega_\mathrm{z}$, we also investigate a simpler quasi-1D model by assuming the atoms to be frozen in the ground state of the radial oscillator potential, resulting in effective quasi-1D GPE
\begin{equation}
i \hbar \frac{\partial \psi}{\partial t} = \left[ \frac{-\hbar^2}{2m} \frac{\partial^2}{\partial z^2} + \frac{m \omega_\mathrm{z}^2 z^2}{2} + V(z) + g_\mathrm{1D} |\psi|^2 \right] \psi, \label{eqn:1dgpe}
\end{equation}
with effective interaction strength
\begin{equation}
g_\mathrm{1D} = 2 \hbar \omega_\mathrm{r} a_\mathrm{s} N.
\end{equation}

In all simulations, we take $a_\mathrm{s} = \SI{-10}{a_0}$ (where $a_0$ is the Bohr radius), $\omega_\mathrm{r} = 2 \pi \times \SI{40}{Hz}$ and $\omega_\mathrm{z} = 2 \pi \times \SI{1.4}{Hz}$. We generally take $w_\mathrm{x} = \SI{117}{\micro m}$, although simulations for the wide barrier (Figs.~\ref{splitting_figure} and \ref{wide_barrier}) have $w_\mathrm{x} = \SI{434}{\micro m}$. We take $N=2500$ in quasi-1D ($N=2000$ in 3D). We assume the initial condition to be a soliton in its ground state with respect to the waveguide potential, displaced to position $-z_0$. The parameters $w_z$, $z_\mathrm{off}$ and $z_0$ are varied as described elsewhere. Our numerical simulations all use Fourier pseudospectral methods and a 4th-5th order adaptive Runge-Kutta scheme to obtain the real-time evolution. Our 3D simulations run on GPUs and obtain ground states using imaginary time evolution. Our quasi-1D simulations obtain ground states using an iterative biconjugate gradient scheme.

\subsubsection*{Approximate analytic quasi-1D model}

To understand the nature of recombination, particularly the appearance of
interference fringes as $z_\mathrm{off}$ is varied, we outline a simple
approximate analytic model for the fraction of the atoms found on the
``right'' of the barrier ($z > z_\mathrm{off}$) following the second barrier
interaction. For convenience, we denote this quantity $A \equiv N_\mathrm{R} /
N_\mathrm{tot}$ and our analytic estimate of it $A_\mathrm{est}$.  It is
convenient to develop the following in dimensionless ``soliton'' units defined
by length $\ell_\mathrm{s} = \hbar / (2 m \omega_\mathrm{r} |a_\mathrm{s}| N)$,
time $t_\mathrm{s} = \hbar/(4m\omega_\mathrm{r}^2|a_\mathrm{s}|^2N^2)$ and
energy $E_\mathrm{s} = 4m \omega_\mathrm{r}^2 |a_\mathrm{s}|^2 N^2 $. In these
units, the quasi-1D GPE becomes
\begin{equation}
i \frac{\partial \tilde{\psi}}{\partial \tilde{t}} 
= \left[ -\frac{1}{2} \frac{\partial^2}{\partial \tilde{z}^2} 
+ \frac{E_\mathrm{h}^2 \tilde{z}^2}{2 E_\mathrm{s}^2} 
+ \frac{\tilde{q}}{\tilde{\ell}} \sqrt{\frac{2}{\pi}} 
    \exp \left( \frac{-2 \tilde{z}^2}{\tilde{\ell}^2} \right)
- |\tilde{\psi}|^2 \right] \tilde{\psi},
\end{equation}
where $\tl = w_\mathrm{z} / \ell_\mathrm{s}$ and
\begin{equation}
\tilde{q} = \frac{\alpha P}{2 \sqrt{2 \pi} \hbar \epsilon_0 c w_\mathrm{x} \omega_\mathrm{r} |a_\mathrm{s}| N}.
\end{equation}
Dimensionless velocities are given by
\begin{equation}
\tilde{\vv} = \frac{\vv}{2\omega_\mathrm{r} |a_\mathrm{s}| N}.
\end{equation}

Ignoring the relative phase of the solitons, the timing of the various barrier interactions, the interatomic interactions, and assuming the waveguide potential to be constant over the width of the barrier, $A_\mathrm{est}$ can be approximated by using the known result for scattering from a $\mathrm{sech}^2$ potential \cite{landau_quantum}{.} Specifically, approximating the optical barrier potential by
\begin{equation}
V(z) = \frac{\tilde{q}}{\tilde{\ell}} \sqrt{\frac{2}{\pi}} 
    \exp \left( \frac{-2 z^2}{\tilde{\ell}^2} \right)
\approx
 \frac{\tilde{q}}{\tilde{\ell}} \sqrt{\frac{2}{\pi}} 
    \sech^2 \left( \frac{\pi z}{\kappa \tilde{\ell}} \right)\,,
\end{equation}
where $\kappa = (\pi/2)^{3/2}$, we obtain the approximate transmission probability \textit{for a single barrier interaction}
\begin{equation}
T(\tv) = \frac{\sinh^2(\kappa \tl \tv)}{\sinh^2(\kappa \tl \tv) 
+ \cosh^2 \left(\frac{\pi}{2} \sqrt{4 \kappa \tq \tl /\pi - 1} \right)}\,.
\end{equation}

We assume that the atoms impact the barrier at the first interaction in the form of an ideal soliton with dimensionless velocity $\tv_\mathrm{in} = z_0 \omega_\mathrm{z} / (2 \omega_\mathrm{r} |a_\mathrm{s}| N)$, chosen such that in the quasi-1D numerics the transmitted fraction is $1/2$. The velocity $\tv_\mathrm{half}$ at which
$T(\tv_\mathrm{half}) = 1/2$ is typically very close to this. After the first barrier interaction, the amplitudes of the outgoing wavepackets in momentum space are approximately
\begin{align}
|\tilde{\psi}_{z>z_\mathrm{off}} (\tv)| &\approx t(\tv - \tv_\mathrm{in} + \tv_\mathrm{half}) \sech[\pi (\tv - \tv_\mathrm{in})] \sqrt{\pi/2},\label{eqn:splitanalytict}\\
|\tilde{\psi}_{z<z_\mathrm{off}} (\tv)| &\approx r(\tv - \tv_\mathrm{in} + \tv_\mathrm{half}) \sech[\pi (\tv - \tv_\mathrm{in})]\sqrt{\pi/2},\label{eqn:splitanalyticr}
\end{align}
where $t(\tv) = T(\tv)^{1/2}$ and $r(\tv) = [1-T(\tv)]^{1/2}$. Between the
first and second barrier interactions the wavepackets undergo nonlinear evolution. We empirically approximate this by assuming that the wavepackets re-form into solitons with half of the initial amplitude, preserving the location in momentum space of the \textit{peak} of the transmitted or reflected amplitude, as obtained from
Eqs.~(\ref{eqn:splitanalytict}--\ref{eqn:splitanalyticr}). This gives wavepackets incoming to the second barrier interaction
\begin{align}
\tilde{\psi}_{z>z_\mathrm{off}}' (\tv) &\approx \sech[2 \pi (\tv - \tv_{\mathrm{t,peak}})]\sqrt{\pi/2},\label{eqn:reformedanalytict}\\
\tilde{\psi}_{z<z_\mathrm{off}}' (\tv) &\approx \sech[2 \pi (\tv - \tv_{\mathrm{r,peak}})]\sqrt{\pi/2},\label{eqn:reformedanalyticr}
\end{align}
where $\tv_{\mathrm{t,peak}}$ and $\tv_{\mathrm{r,peak}}$ denote the numerically-determined locations of the peaks described. Finally, the fraction of the total population to the right is approximated by
\begin{equation}
A_\mathrm{est}^{(0)} = \int \left| t(\tv - \tv_\mathrm{in} + \tv_\mathrm{half})\, \psi_{z<z_\mathrm{off}}' (\tv) + r(\tv - \tv_\mathrm{in} + \tv_\mathrm{half})\, \psi_{z>z_\mathrm{off}}' (\tv) \right|^2 dz.    
\end{equation}

The estimate above will be modulated by interference between the solitons. We model
this using the approach for $\delta$-function barriers described in Ref.~\cite{soliton_collisions_barrier}{.} This is expected to be a good approximation for narrow barriers when the dimensionless soliton velocity at the barrier $\tilde{v}_\mathrm{in} \gtrsim 1$.
Assuming $z_0 \gg z_\mathrm{off}$ the dimensionless velocity of
all solitons is approximately $\pm \tv_\mathrm{in}$ when they are in the neighbourhood of the barrier. In this approximation the two solitons to the right of the barrier are separated by $4 z_\mathrm{off}$. Thus, the wavefunction to the right ($z > z_\mathrm{off}$) immediately after
recombination can be written as
\begin{equation}
\tilde{\psi}_\mathrm{right} (\tilde{z}) \approx \frac{\sqrt{A_\mathrm{est}^{(0)}}}{4} \left[ e^{i \tv_\mathrm{in} \tilde{z}} \sech \left( \frac{\tilde{z} - \tv_\mathrm{in} \tilde{t}}{4} \right) + e^{i \tv_\mathrm{in} (\tilde{z} - 4 \tilde{z}_\mathrm{off})} \sech \left( \frac{\tilde{z} - 4 \tilde{z}_\mathrm{off} - \tv_\mathrm{in} \tilde{t}}{4} \right)  \right],    
\end{equation}
where we have assumed equal-amplitude solitons.
Note that  we have omitted numerous irrelevant phase factors compared to the expressions in Ref.~\cite{soliton_collisions_barrier} for simplicity, and the phase shift of $\pi/2$ gained by the soliton transmitted at the second barrier interaction simply cancels the phase that the soliton reflected at the second barrier interaction previously acquired by being transmitted at the first barrier interaction. Integrating $|\tilde{\psi}_\mathrm{right}(\tilde{z})|^2$, we obtain
\begin{equation}
A_\mathrm{est} = A_\mathrm{est}^{(0)} \left[ 1 + \cos \left( 4 \tv \tilde{z}_\mathrm{off} \right)  \frac{\tilde{z}_\mathrm{off}}{\sinh \left( \tilde{z}_\mathrm{off} \right)} \right] = A_\mathrm{est}^{(0)} \left[ 1 + \cos \left( \frac{4 z_0 z_\mathrm{off}}{\ell_0^2} \right) \frac{z_\mathrm{off} / \ell_\mathrm{s}}{\sinh \left( z_\mathrm{off} / \ell_\mathrm{s} \right)}\right] ,    
\end{equation}
where the final expression is in real units, the velocity $\vv_\mathrm{in}$ has been replaced with $z_0 \omega_\mathrm{z}$, and $\ell_0 = (\hbar / m \omega_\mathrm{z})^{1/2}$ is the axial harmonic oscillator length.

As shown in the Extended Data Figs.~\ref{fig:varywidth} and \ref{fig:varyinitialdisp}, the analytic approximation gives a good qualitative picture of the behavior across a wide regime of parameters and provides a quantitatively accurate value for the fringe spacing across a considerable fraction of this. In particular, we find it can give useful results for barrier interactions with $\tilde{v}_\mathrm{in} \lesssim 1$. However, it does break down for wider barriers and slower barrier interactions.

\bibliographystyle{naturemag}
\bibliography{soliton_splitting_and_recombination}

\noindent\textbf{Acknowledgements}
We thank A. Mackellar for his assistance in producing the 3D rendering in Fig.~\ref{schematic}. This research made use of the Rocket High Performance Computing service at Newcastle University. We acknowledge the UK Engineering and Physical Sciences Research Council (Grants No. EP/L010844/1 and No. EP/K030558/1) for funding. 

\noindent\textbf{Data availability}
The data presented in this letter are available at \href{http://dx.doi.org/doi:10.15128/r1sf2685090}{doi:10.15128/r1sf2685090}.

\onecolumn
\begin{center}
\section*{Extended data}
\end{center}

\setcounter{figure}{0}    

\begin{figure*}[htbp]
    \centering
    \def\svgwidth{0.35\linewidth}
    \graphicspath{{Figures/}}
    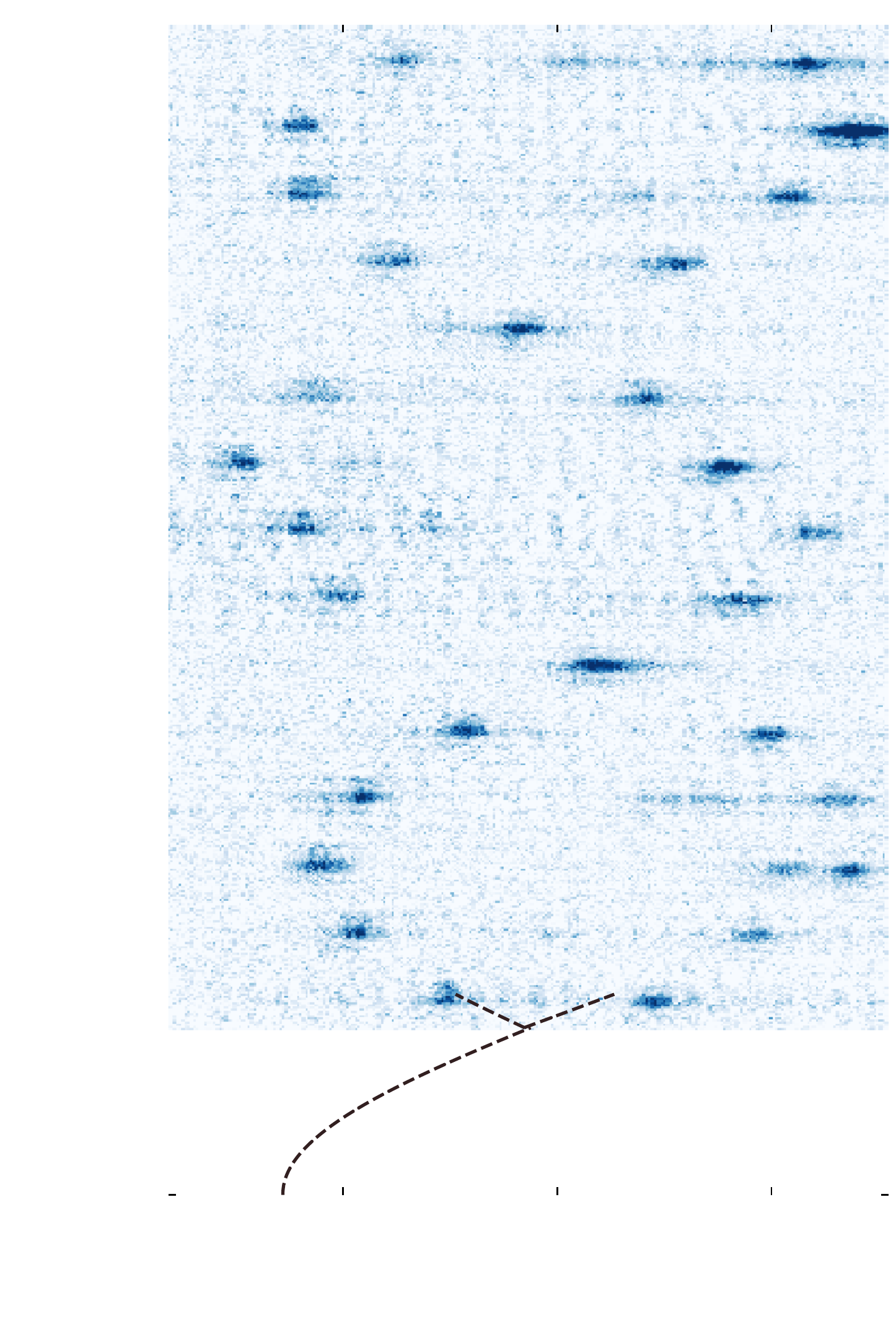
    \caption{Velocity selection during the splitting of a soliton. The initial soliton has $\sim2500$ atoms at a scattering length of $\sim\SI{-8.5}{a_0}$. The transmitted daughter soliton (solid blue trajectory) leaves the wide barrier (green line) with a higher kinetic energy than the reflected soliton (dotted red trajectory) and so oscillates with a clearly larger amplitude. The barrier height is set to give $\sim\SI{50}{\%}$ transmission and is removed after the splitting to allow the solitons to oscillate. The barrier is offset from the centre of the $\SI{0.5}{Hz}$ trap by $\sim\SI{30}{\micro m}$ in this case, although this is not expected to play a significant role in the observed velocity filtering. The black dashed line is the expected trajectory leading up to splitting. This gives an initial release position $\sim\SI{250}{\micro m}$ from the centre of the harmonic potential.}
    \label{velocity_selection}
\end{figure*}
    
\begin{figure*}[htbp]
    \centering
    \def\svgwidth{0.6\linewidth}
    \graphicspath{{Figures/}}
    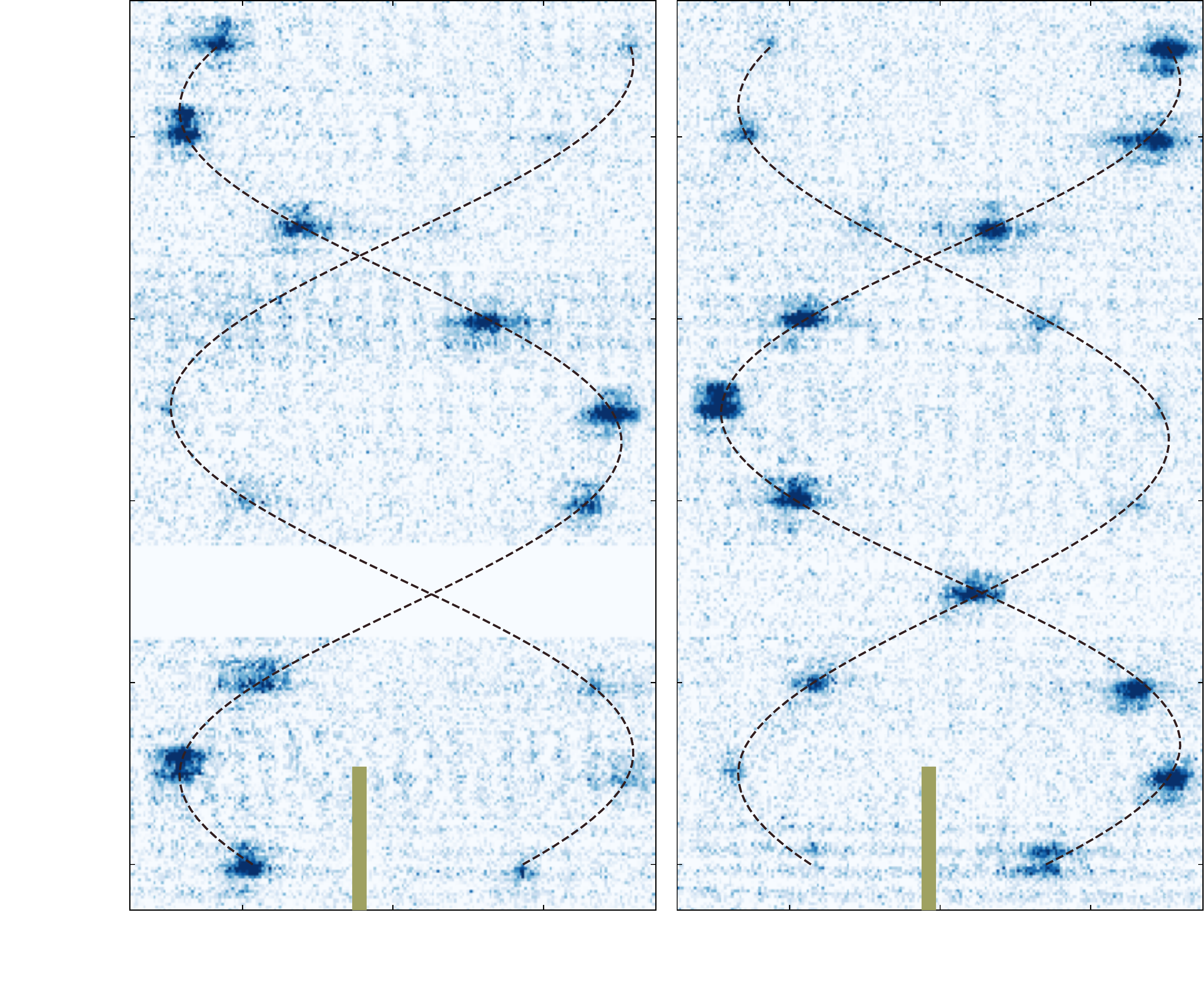
    \caption{Image sequences showing that the two daughter solitons simply pass through one another in the absence of the barrier. The initial soliton is split by the wide barrier into two daughter solitons with asymmetric atom number, allowing us to track their individual trajectories. In [a] the barrier was calibrated to give $\sim\SI{20}{\%}$ transmission and in [b] the barrier was calibrated to give $\sim\SI{80}{\%}$ transmission. As in Ref.~\cite{soliton_collisions_hulet}{,} we cannot disregard the possibility that atom exchange happens during the soliton-soliton collisions. However, these plots show the importance of the barrier for facilitating interference. The image for $\SI{0.55}{s}$ in [a] was not taken and so a space is left blank. The dashed black lines are sinusoidal fits to the soliton trajectories.}
    \label{asymmetric_splitting}
\end{figure*}
    
\begin{figure*}[htbp]
    \centering
    \def\svgwidth{0.8\linewidth}
    \graphicspath{{Figures/}}
    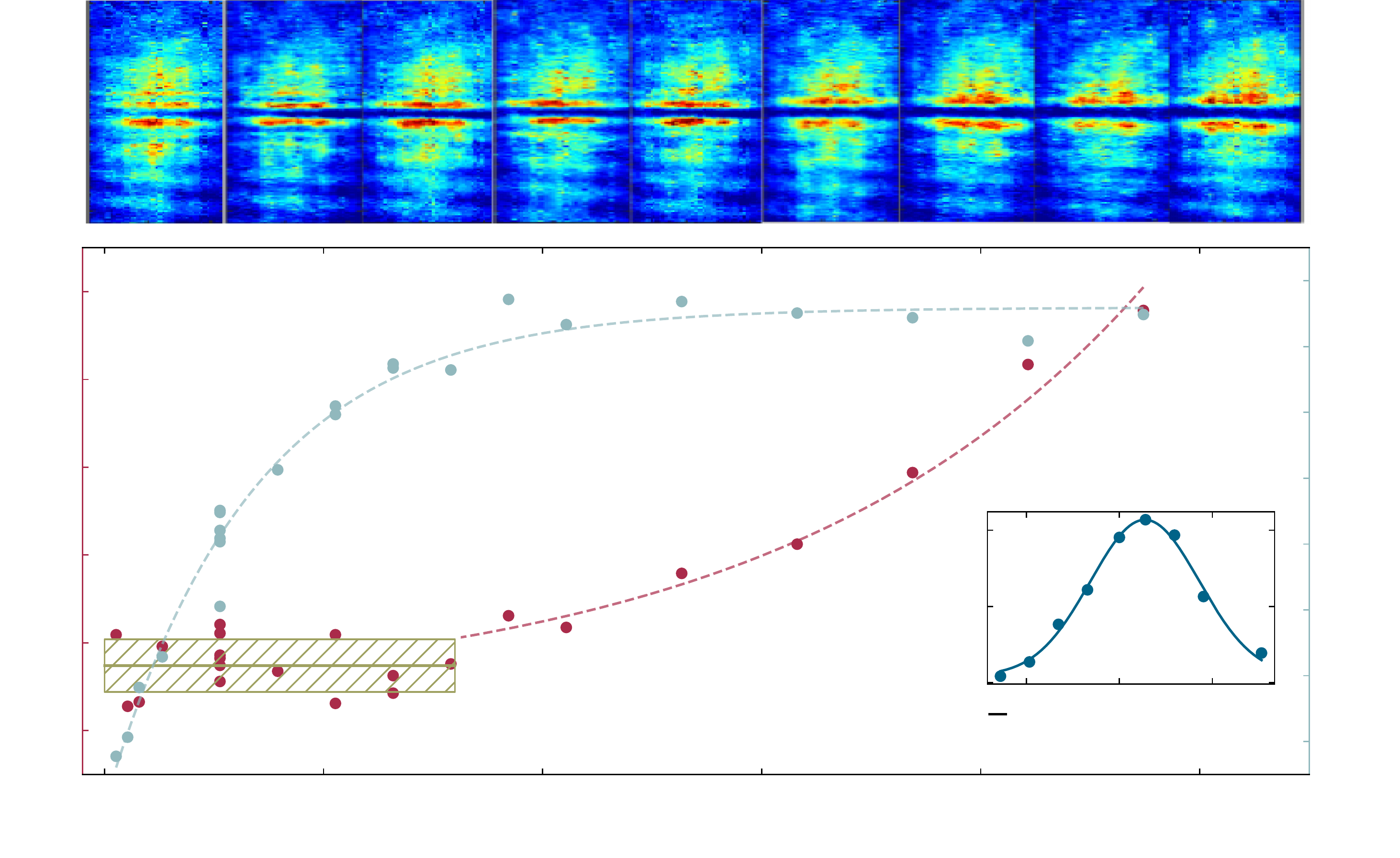
    \caption{Measuring the width of the narrow barrier by depleting part of a thermal atomic cloud. The cloud is exposed to the barrier beam for $\SI{1}{ms}$ before being imaged in-trap. [a] At sufficiently low barrier powers ($P<\SI{8}{mW}$ in this case), the depth of the depleted region varies with power, but the measured width remains constant. This is because the imprint left by the barrier beam is not totally depleting the atoms from the region of interest. The solid green line and hatched area show the mean barrier width and $\pm \SI{1}{s{.}d{.}}$ for all data in the partial depletion region. The dashed lines are guides to the eye. [b] Example images of atom depletion for a range of barrier powers from $\SI{2.64}{mW}$ to $\SI{23.72}{mW}$ (left to right, respectively). The additional fringes are imaging artefacts and are not included in the fits of the width. [c] Using an acousto-optic deflector to steer the waveguide, we move the atomic cloud across the transverse direction of the barrier to determine its transverse width.}
    \label{cloud_depletion}
\end{figure*}

\begin{figure*}[htbp]
    \centering
    \includegraphics[width=0.5\columnwidth]{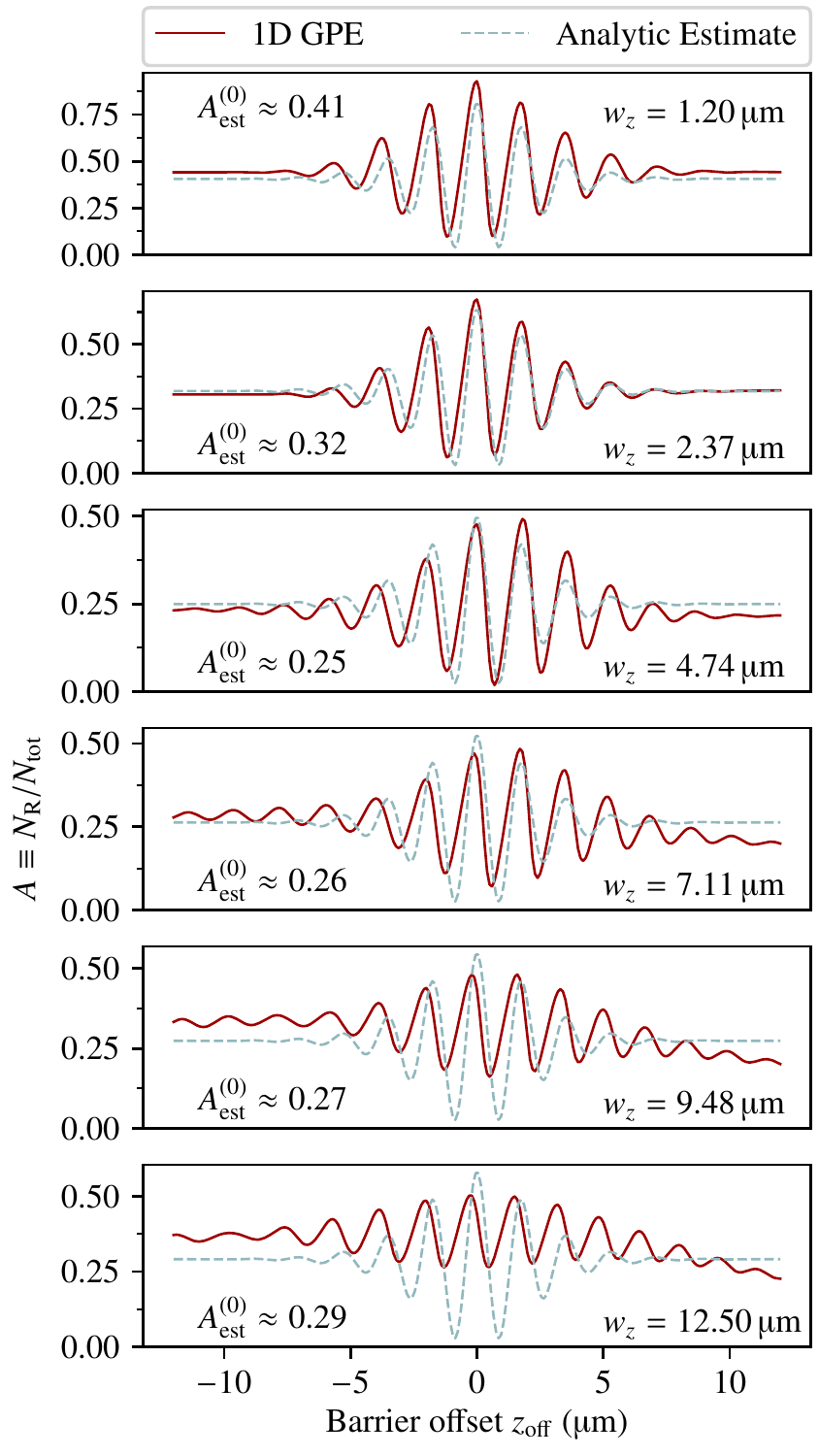}
    \caption{Theoretical modelling: recombined fraction to right of barrier as a function of barrier offset, showing variation as a function of barrier width.}
    \label{fig:varywidth}
\end{figure*}

\begin{figure*}[htbp]
    \centering
    \includegraphics[width=0.5\columnwidth]{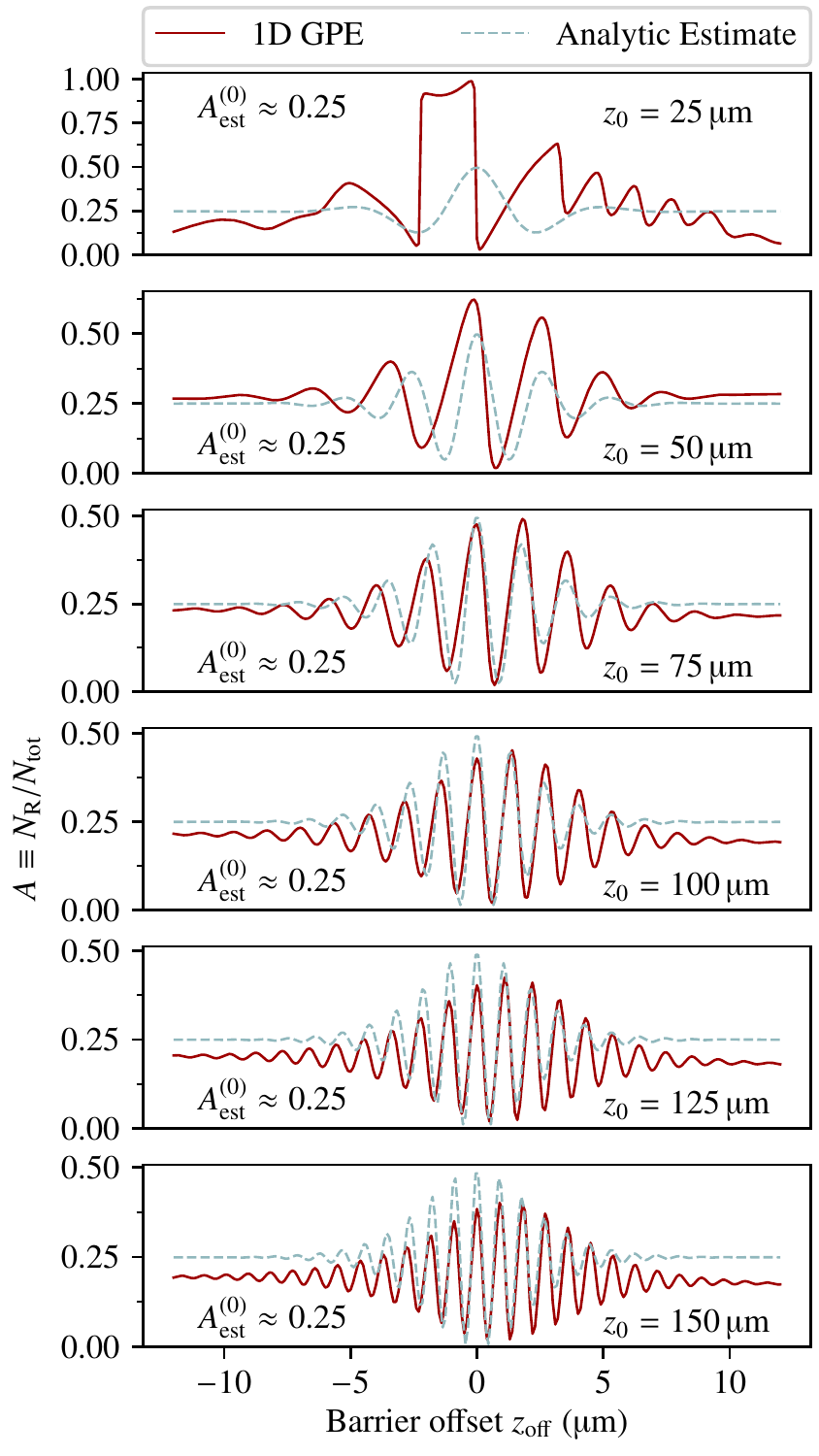}
    \caption{Theoretical modelling: recombined fraction to right of barrier as a function of barrier offset, showing variation as a function of initial soliton displacement.}
    \label{fig:varyinitialdisp}
\end{figure*}

\begin{figure*}[htbp]
    \centering
    \def\svgwidth{0.9\linewidth}
    \graphicspath{{Figures/}}
    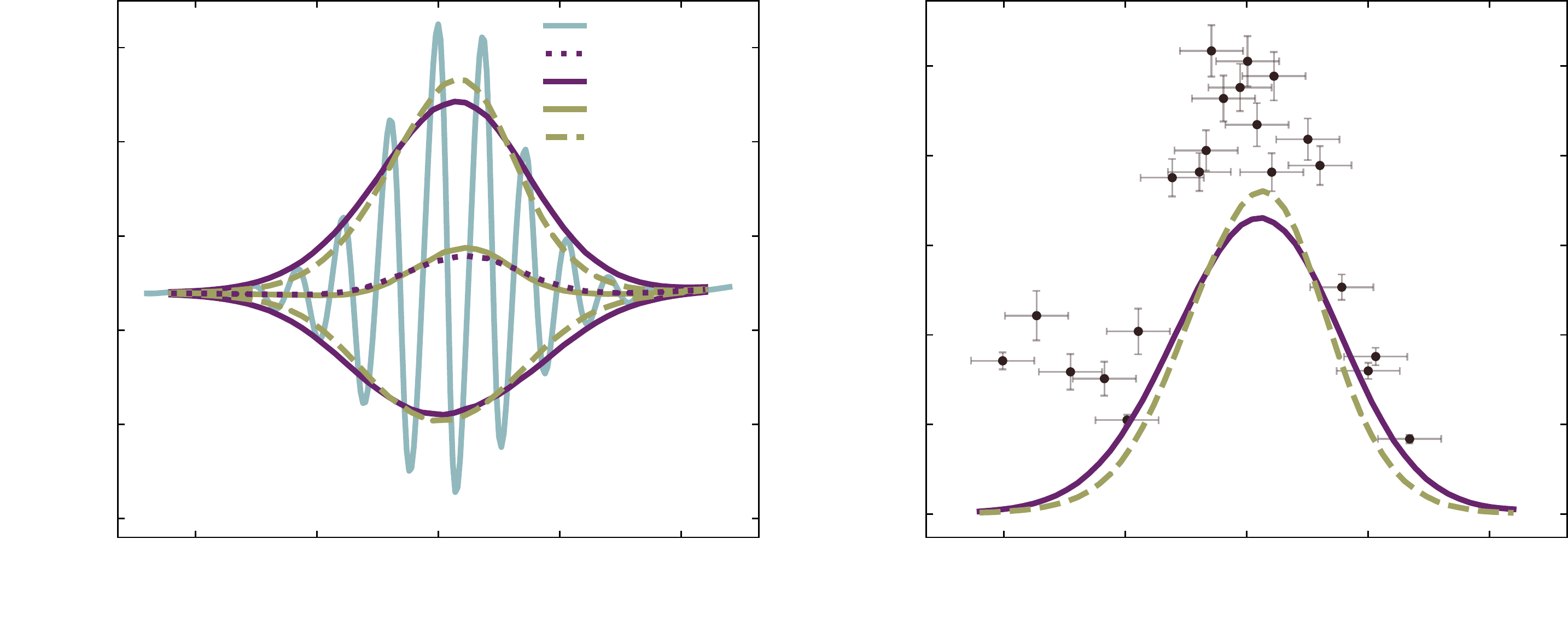
    \caption{Comparing phase noise and barrier offset noise as the source of shot-to-shot fluctuations observed in the experiments with the narrow barrier. [a] $1$D-GPE simulations predict oscillatory behaviour when the barrier is offset, as seen in Fig.~\ref{recombination} of the main text. A sampling of these fringes is performed by interpolation to simulate the $\SI{1.3}{\micro m}$ uncertainty in barrier offset. From this, we determine a mean (labelled `$z_\mathrm{off}$, mean') and an envelope defined by the standard deviation (labelled `$z_\mathrm{off}$, SD'). We achieve a similar mean and envelope if we instead ascribe a random phase to each daughter soliton following splitting (labelled `$\phi$, mean' and `$\phi$, SD' respectively). In [b] we compare the standard deviation of the experimental data (black) with that of the random offset and random phase models (purple solid line and green dashed line respectively). The vertical error bars on the data are estimates of the fractional uncertainties in the error \cite{hughes2010measurements}{.} While the detailed origin of the discrepancy between the experimental the theoretically modelled data is not known, they do clearly agree qualitatively on the existence and location of the peak.}
    \label{standard_deviation}
\end{figure*}

\end{document}